\documentclass[12pt]{article}

\begin{document}

%\addtolength{\textheight}{4cm}

%\setlength{\baselineskip}{0.33in}

\begin{flushright}
IMSc/2007/09/12
\end{flushright}

\vspace{2mm}

\vspace{2ex}

\begin{center}
{\large \bf A Description of Rotating Multicharged Black Holes} \\

\vspace{2ex}

{\large \bf in terms of Branes and Antibranes} \\

\vspace{8ex}

{\large  Samrat Bhowmick and S. Kalyana Rama}

\vspace{3ex}

Institute of Mathematical Sciences, C. I. T. Campus, 

Tharamani, CHENNAI 600 113, India. 

\vspace{1ex}

email: samrat, krama@imsc.res.in \\ 

\end{center}

\vspace{6ex}

\centerline{ABSTRACT} 
\begin{quote}

We describe rotating multicharge black holes as stacks of
intersecting branes and antibranes together with massless gases
on them. Assuming the energies of the gases to be equal, we find
that their angular momentum parameters, corresponding to black
hole rotations, are also equal. The entropy $S$ of this model is
given by $S = X S_{sg}$ where $S_{sg}$ is the supergravity
entropy. One can obtain $X = 1$ under an assumption which
violates conservation of energy. We show that $X = 1$ can also
be obtained if one assumes that there is only one single gas,
which is some sort of superposition of the gases mentioned
above, and that the brane tensions are reduced by a factor of
four. In this interpretation, energy is conserved and the
unusual assumption that energies, not temperatures, of the gases
are equal becomes superfluous.

\end{quote}

\vspace{2ex}

%PACS numbers: 98.80.Cq

\newpage

\section{Introduction}

Enormous progress has been made in understanding the microscopic
origin of entropy and Hawking radiation of extremal and near
extremal black holes using various brane configurations in
string or M theory and low energy excitations on them
\cite{bhrev}. Despite a variety of attempts \cite{Sbh}, a
similar level of understanding of non extremal black holes, in
particular Schwarzschild black holes, is lacking.

A few years ago, Danielsson, G\"{u}ijosa and Kruczenski (DGK)
proposed a field theoretic model for non extremal black holes
\cite{DGK}. Their description is valid far from extremality,
thus also for Schwarzschild black holes. They considered stacks
of D3, M2, or M5 branes and antibranes together with massless
gases on them. This field theoretic model has been generalised
to other single charge black holes with or without rotation
\cite{ght, ser-peet, bl}, described as stacks of Dp branes and
antibranes together with massless gases on them; and, also to
multicharge black holes with no rotation \cite{krama, lif, ks},
described as stacks of intersecting branes and antibranes
\cite{ib} together with $2^K$ types of massless gases on them
where $K$ is the number of charges.

The gases are characterised by their energies $E_1, E_2, E_3,
\cdots, E_{2^K} \;$. Their entropies are obtained from the near
extremal limit of the corresponding supergravity solutions.
Assuming the gases to have equal energies, instead of equal
temperatures which is usually the case, and following the
analysis of \cite{DGK}, it has been found that non rotating
black hole entropy $S$ in the field theoretic model is
identical, upto a constant numerical factor, to the entropy
$S_{sg}$ in the supergravity description. That is, $S = X
S_{sg}$ where the `deficit factor' $X$ is a constant that
depends on $K$ and the number of transverse dimensions.

Furthermore, as noted in \cite{DGK}, one can obtain $X = 1$ for
non rotating black holes under a further assumption about the
energies of the gases which, however, violates conservation of
energy. This violation, as noted in \cite{ser-peet}, is perhaps
due to the binding energy of branes and antibranes not being
taken into account. Alternately, one can obtain $X = 1$ with no
violation of conservation of energy if, as shown in \cite{krama,
ks}, one assumes that the available energy is all taken by one
single gas, which is some sort of superposition of $2^K$
possible types of gases, and further assumes that its entropy is
an average of the $2^K$ gas entropies and that all the brane
tensions are reduced by a factor of four. The unusual assumption
that energies, not temperatures, of the gases are equal then
becomes superfluous.

Consider black holes with rotation. In the case of single charge
rotating black holes, the gases are also characterised by
angular momentum parameters $l_j$ where the subscript $j$
denotes possible rotations in the transverse space. Assuming the
energy and the parameters $l_j$ to be the same for the gases on
branes and antibranes, and with a further assumption about the
energies of the gases which violates conservation of energy, it
is shown in \cite{ght, ser-peet} that the field theoretic and
supergravity entropies agree, namely $S = S_{sg} \;$. However,
without the last assumption above, one obtains $S = X S_{sg}$
where the deficit factor $X$ is not constant but depends on
black hole parameters; it becomes constant when rotation is
absent.

In this paper we study rotating multicharge black holes,
described as stacks of intersecting branes and antibranes
together with $2^K$ types of gases. The gases are also
characterised by angular momentum parameters which correspond to
black hole rotations and are, in general, different for each
gas.

We assume that the energies of the gases are equal and that the
angular momentum parameters are different. Following the methods
of \cite{DGK, ght, ser-peet}, we then find that the angular
momentum parameters are also equal for all the $2^K$ types of
gases. We also find that the field theoretic and supergravity
entropies agree, namely $S = S_{sg}$, under a further assumption
about the energies of the gases which violates conservation of
energy. Without this assumption, one obtains $S = X S_{sg}$
where the deficit factor $X$ is not constant but depends on
black hole parameters; it becomes constant when rotation is
absent. These results are similar to those for single charge
rotating black holes \cite{ght, ser-peet}, but are now shown to
be valid for multicharge black holes also.

The result that the deficit factor $X$ is not constant when
rotation is present makes the field theoretic model less
appealing. The assumption under which one obtains $X = 1$ is not
satisfactory since it violates conservation of energy. This
violation is perhaps due to the neglect of binding energies
\cite{ser-peet}, but the details of the binding energies are not
sufficiently well known to verify this idea. Note also the
necessity of the unusual assumption that energies, not
temperatures, of the gases are equal in obtaining this result.

In contrast, the deficit factor $X = 1$ was obtained in
\cite{krama, ks} for non rotating multicharge black holes
without the assumptions mentioned above. In this paper, we show
that $X = 1$ can be obtained similarly for rotating multicharge
black holes also.

This paper is arranged as follows. In section {\bf 2} we give
relevant results for rotating multicharge black holes from
supergravity description. In sections {\bf 3.1} and {\bf 3.2} we
describe the field theoretic model for rotating single and
multicharge black holes. In section {\bf 4} we show that $X = 1$
can be obtained without the assumptions mentioned earlier. In
section {\bf 5} we conclude with a brief summary and a few
issues for further study.

\section{Supergravity description of \\
multicharge rotating black holes}

In string theory, spinning p-branes describe rotating black
holes. Spinning intersecting p-branes describe multicharge
rotating black holes. Supergravity solutions for various
intersecting branes have been studied \cite{bhrev}. An algorithm
for finding explicit solutions for multicharge black holes is
given in \cite{cy1}.

Supergravity expressions for mass, angular momenta, and entropy
of single charge rotating black holes may be found in \cite{ho}.
For two charge cases, it may be found in \cite{cy2}. In general,
the supergravity expressions for mass $M_{sg} \;$, angular
momenta $J_{j \; sg} \;$, and entropy $S_{sg} \;$, of K --
charge rotating black holes can be written as
\begin{eqnarray}
M_{sg}& = & b \left( 2 \lambda \mu 
+ \sum_{i=1}^{K} \sqrt{Q_{i \; sg}^2 + \mu^2} \right) \\
J_{j \; sg} & = & \frac{2 b}{n} \; l_{j \; sg} \; 
(2 \mu)^{ \frac{2 - K}
{2}} \prod_{i=1}^{K} \left( \sqrt{\mu^2 + Q_{i \; sg}^2} + \mu
\right)^{\frac{1}{2}} \\
S_{sg} & = & \frac{4 \pi b}{n} \; 
r_{H_{sg}} \; (2 \mu)^{\frac{2 - K}
{2}} \prod_{i=1}^{K} \left( \sqrt{\mu^2 + Q_{i \; sg}^2} + \mu
\right)^{\frac{1}{2}}
\end{eqnarray} 
where $r_{H_{sg}}$ is the radius of horizon which is given by
the equation
\begin{equation}\label{sugra_r} 
r_{H_{sg}}^n \prod_j \left( 1 + \frac{l_{j \; sg}^2}
{r_{H_{sg}}^2} \right) = 2 \mu \; .
\end{equation} 
In the above expressions, $\mu$ is a non extremality parameter,
$Q_{i \; sg}$ are the charges, $l_{j \; sg}$ are the rotation
parameters, $\lambda = \frac{n+1}{n} - \frac{K} {2}$, and $b =
\frac{n \omega_{n+1} V_p} {(2 \pi)^7 g_s^2 l_s^8}$ where
$\omega_{n + 1}$ is the area of an unit $(n + 1)$ -- dimensional
sphere, $V_p$ is the volume of the $p$ -- dimensional compact
space, and $n = 7 - p$. Here and in the following, the
subscripts $i = 1, 2, \cdots, K$ always refer to the charges and
the subscripts $j = 1, 2, \cdots, \left[ \frac{n + 2}{2}
\right]$ always refer to the rotations in the $(n + 3)$ --
dimensional transverse space.

Note that $\frac{2\pi J_{j \; sg}}{S_{sg}} = \frac{l_{j \; sg}}
{r_{H_{sg}}}$. Define $\rho_{sg}$ by the equation
\[
\rho_{sg}^n = \frac{r^n_{H_{sg}}}{2 \mu} = 
\prod_j \left( 1 + \frac{l_{j \; sg}^2}{r_{H_{sg}}^2} 
\right)^{ - 1} \; . 
\]
Then, $S_{sg}$ can be written as 
\[
S_{sg} = \rho_{sg} \; S_{0 \; sg} \;,\;\;\;\; S_{0 \; sg} 
= \frac{4
\pi b}{n} (2 \mu)^\lambda \prod_{i=1}^{K}\left( \sqrt{\mu^2 +
Q_{i \; sg}^2} + \mu \right)^{\frac{1}{2}}
\]
where $S_{0  \; sg}$ is the entropy in the non rotating case, namely
when $l_{j \; sg} = 0$. 

The charges $Q_{i \; sg}$ can also be parametrised in terms of
$\phi_i$ as follows :
\begin{equation}
Q_{i \; sg} = \mu \; \sinh{2 \phi_i} \; \; , \; \; \; 
i = 1, 2, \cdots, K \; . 
\end{equation}
The mass, angular momenta, and entropy are then given by
\begin{eqnarray}
M_{sg} & = & b \mu \; \left( 2 \lambda 
+ \sum_{i=1}^K \cosh{2 \phi_i} \right) \\
J_{j \; sg} & = & \frac{4 b \mu}{n} \; l_{j \; sg} \; 
\prod_{i=1}^{K} \cosh \phi_i \\
S_{sg} & = & \frac{8 \pi b \mu}{n} \; r_{H_{sg}} \; 
\prod_{i=1}^{K} \cosh \phi_i \; \; . 
\end{eqnarray}

In the no-rotation and extremal limit, where $l_{j \; sg} = 0$,
$\mu \to 0$ and $\phi_i \to \infty$ such that $Q_{i \; sg}$
remain finite, the mass becomes $M_{sg \; ext} = \sum_i b Q_{i
\; sg} \;$. In string theory, non rotating extremal black holes
are identical to stacks of intersecting p--branes. Let such a
stack consist of $N_i$ number of $i^{th}$ type of branes with
tension $\tau_i$ and volume $V_i$. The total brane mass is then
$M = \sum_i N_i \tau_i V_i \;$. Identifying these two masses
then gives the relation between the charges $Q_{i \; sg}$ and
the number $N_i$ of $i^{th}$ type of branes: $N_i = \frac{b Q_{i
\; sg}}{\tau_i V_i} \;$.

A little calculation enables one to get near extremal values of
mass, angular momenta, and entropy. To the leading order in
$\mu$, they are given by
\begin{eqnarray}
M_{sg} & = & \sum_{i = 1}^{K} b Q_{i \; sg} + E \label{Mne} \\
J_{j \; sg} & = & \frac{2 b}{n} \; l_{j \; sg} \; 
\left( \frac{E}{\lambda b} \right)^{\frac{2 - K}{2}} 
\prod_i \sqrt{Q_{i \; sg}} \label{Jjne} \\
S_{sg} & = & \frac{4\pi b}{n} \; r_{H_{sg}} \; 
\left( \frac{E}{\lambda b} \right)^{\frac{2 - K}{2}}
\prod_i \sqrt{Q_{i \; sg}} \label{Sne}
\end{eqnarray}
where $E = 2 b \lambda \mu $ is the energy above extremality, and
$r_{H_{sg}}$ is given by
\begin{eqnarray}
r_{H_{sg}}^n \prod \left( 1 + \frac{l_{j \; sg}^2}
{r_{H_{sg}}^2}
\right) = \frac{E}{\lambda b} \; \; . \label{rHne}
\end{eqnarray}

\section{Field theoretic description of \\
rotating black holes }

\subsection{Single charge rotating black holes}

Description of single charge rotating black holes in the field
theoretic model was given in \cite{ght, ser-peet}. Following
them we consider, as in non rotating case \cite{DGK}, stacks of
branes and antibranes together with a massless gas on each
stack. So the model consists of (i) $N$ spinning branes, (ii)
$\bar{N}$ spinning antibranes, and (iii) two gases of massless
excitations on branes and antibranes. The masses of branes and
antibranes are $\tau V_p N$ and $\tau V_p \bar{N}$, and their
charges $q$ and $\bar{q}$ are $\frac{\tau V_p N}{b}$ and
$\frac{\tau V_p \bar{N}}{b} \;$, $\tau$ being brane tension and
$V_p$ being brane volume.

Upto now we considered a system very similar to one considered
in \cite{DGK}. Together with the above set up, this time we also
consider two sets of extra parameters $l_j$ and $\bar{l}_j$
which are the charges corresponding to the symmetries of the
rotational space.  These `charges', which are analogous to
R-symmetry charges in the case of D3 branes, characterise
rotation of black hole. We will write these charges as angular
momentum parameters.

This system is very similar to one given in \cite{ght,
ser-peet}. Unlike in these works, however, we assume here that
different gases have different angular momentum parameters, {\em
i.e.} that $l_j$ and $\bar{l}_j$ are different. Here, in the
field theoretic model, near extremal non rotating black holes
are described as stacks of branes and antibranes. We assume that
branes and antibranes do not interact with each other. The total
mass, angular momenta, and entropy will then be additive. Also,
these quantities are assumed to be given, as in \cite{DGK}, by
the supergravity expressions in the near extremal regime, namely
by equations (\ref{Mne}) -- (\ref{rHne}).

Moreover, the energies of the gas on the branes and antibranes
are assumed to be equal, namely $E = \bar{E}$. Note that,
normally, subsystems of any given system all have the same
temperature. Hence, normally, temperature of branes and
antibranes should have been set equal. But here, instead of
temperature, one has to assume that energies are equal;
otherwise field theoretic and the supergravity entropies do not
match. For now, we make this assumption and continue our
analysis, although no physical mechanism is known which enforces
equality of energies, instead of temperatures, among the
subsystems. Later in the paper, we will see how this assumption
becomes superfluous.

With $E = \bar{E}$, the total mass $M$, charge $Q$, angular
momenta $J_j$, and entropy $S$ in the field theoretic model here
will be
\begin{eqnarray}
\label{M_ft} 
M & = & b (q + \bar{q}) + 2 E \\ 
\label{Q_ft}
Q & = & q - \bar{q} \\
\label{J_ft}
J_j & = & \frac{2b}{n \sqrt{b \lambda}} \left( l_j \sqrt{q} 
+ \bar{l}_j \sqrt{\bar{q}} \right) \sqrt{E} \\
\label{S_ft}
S & = & \frac{4 \pi b}{n \sqrt{b \lambda}} \left(r_H \sqrt{q} 
+ \bar{r}_H \sqrt{\bar{q}} \right) \sqrt{E} 
\end{eqnarray}
where $\lambda = \frac{n + 1}{n} - \frac{K}{2} = \frac{n + 2}{2
n}$ for the single charge case $K = 1$. In the above
expressions, $r_H$ and $\bar{r}_H$ are functions of
$(E,\{l_j\})$ and $(E,\{\bar{l}_j\})$ respectively and are given
by
\begin{equation}\label{r_H_ft}
r_H^n \; \prod_j \left( 1 + \frac{l_j^2}{r_H^2} \right) 
\; = \bar{r}_H^n \; \prod_j \left( 1 + \frac{\bar{l}_j^2}
{\bar{r}_H^2} \right) = \frac{E}{\lambda b} \; \; .
\end{equation}

This is our field theoretic model for single charge rotating
black holes, with the entropy $S$ given by equation
(\ref{S_ft}). The parameters $q$, $\bar{q}$, $l_j$, $\bar{l}_j$,
and $E$ are arbitrary subject only to the constraints that the
quantities $M$, $Q$ and $J_j$ are fixed.

This system is dynamical because more and more brane antibrane
pairs can be created taking energy from massless gases, or
annihilated giving energy to them; moreover $l_j$ and
$\bar{l}_j$ can flow from one gas to another. The system reaches
equilibrium in a state where entropy is maximum for given $M$,
$Q$, and $J_j$. So, to obtain the equilibrium state, we maximise
$S$ subject to the constraints that $M$, $Q$ and $J_j$ given in
equations (\ref{M_ft}) -- (\ref{J_ft}) are held fixed.

These constraints are incorporated by Lagrange multiplier
method. Hence we maximise the function $\mathcal{F} (q, \bar{q},
l_j, \bar{l}_j, E)$ defined by
\begin{eqnarray*}
\mathcal{F} & = & S + A_j\left\{\frac{2b} 
{n \sqrt{b \lambda}} \left( l_j \sqrt{q} + \bar{l}_j
\sqrt{\bar{q}} \right) \sqrt{E} - J_j \right\} \\
& & +  
\left\{ B (q - \bar{q} - Q) 
+ C \left( b (q + \bar{q}) + 2 E - M \right) \right\}
\end{eqnarray*}
where $A_j$, $B$ and $C$ are Lagrange multipliers. Varying
$\mathcal{F}$ with respect to $q$, $\bar{q}$, $l_j$,
$\bar{l}_j$, and $E$, we have that $d \mathcal{F} = 0$ at the
maximum. Hence,
\begin{eqnarray*}
& & 0 = 
2 \; \left\{ \frac{\partial r_H} {\partial l_j} + \frac{A_j}{2 \pi} \right\}
\sqrt{q E} \;\; d l_j
+ 2 \; \left\{ \frac{\partial \bar{r}_H} {\partial \bar{l}_j} 
+ \frac{A_j}{2 \pi} \right\} \sqrt{\bar{q} E} \;\; d \bar{l}_j \\
& & + \left( \frac{\partial S}{\partial q} + \frac{b}{n\sqrt{b \lambda }}A_j l_j \sqrt{\frac{E} {q}} + B + b C
\right) d q \\ 
& & + \left( \frac{\partial S}{\partial \bar{q}} + \frac{b}{n\sqrt{b \lambda }}A_j \bar{l}_j \sqrt{\frac{E} {q}} - B + b C
\right) d \bar{q} \\
& & + \left\{
\frac{\partial S}{\partial E}+\frac{A_j b}{n \sqrt{b \lambda}}(l_j \sqrt{q}+\bar{l}_j \sqrt{\bar{q}}) \frac{1}{\sqrt{E}} + 2C
\right\} \; dE
\end{eqnarray*}
which implies that the coefficients of $dl_j$, $d\bar{l}_j$,
$dq$, $d\bar{q}$ and $dE$ must vanish. Equating the coefficients
of $dl_j$ and $d\bar{l}_j$ to zero, we get
\begin{equation}\label{max:r_H}
\frac{\partial}{\partial l_j} \; r_H =
\frac{\partial}{\partial\bar{l}_j} \; \bar{r}_H \; .
\end{equation}
Using equations (\ref{r_H_ft}) for $r_H$ and $\bar{r}_H$, one
gets
\begin{eqnarray}
\frac{\partial r_H}{\partial l_k} & = & 
- \; \frac{2 l_k r_H} {r_H^2 + l_k^2}
\left[ n - \sum_j \frac{l_j^2} {r_H^2 + l_j^2} \right]^{-1} 
\label{dr/dl} \\
\frac{\partial \bar{r}_H}{\partial \bar{l}_k} & = & 
- \; \frac{2 \bar{l}_k \bar{r}_H} {\bar{r}_H^2 + \bar{l}_k^2}
\left[ n - \sum_j \frac{\bar{l}_j^2} {\bar{r}_H^2 + \bar{l}_j^2}
\right]^{-1} \; \; .  \label{bar:dr/dl}
\end{eqnarray}
In the above equations, $r_H$ and $\bar{r}_H$ are functions of
$l_j$ and $\bar{l}_j$ respectively. Substituting them in
equation (\ref{max:r_H}) then leads to a complicated
equation. Equation (\ref{max:r_H}), however, always admits one
solution given by $l_j = \bar{l}_j$ and, hence, $r_H =
\bar{r}_H$. In the following, we take this to be the solution,
which is always present. Also, as will be seen below, this
solution leads to the correct form of entropy.

With $l_j = \bar{l}_j$ and $r_H = \bar{r}_H$, the angular
momenta $J_j$ and entropy $S$ now become
\begin{eqnarray}
J_j &=& \frac{2 b}{n \sqrt{b \lambda}} \; 
l_j \; (\sqrt{q} + \sqrt{\bar{q}}) \sqrt{E} \\
S &=& \frac{4\pi b}{n \sqrt{b \lambda}} \; 
r_H \; (\sqrt{q} + \sqrt{\bar{q}}) \sqrt{E} \; . 
\end{eqnarray}
Note that we now get $\frac{l_j}{r_H} = \frac{2 \pi J_j} {S}$
since $l_j = \bar{l}_j$ and $r_H = \bar{r}_H$. Also, just as in
the supergravity case, define $\rho$ by the equation
\begin{eqnarray}\label{rho}
\rho^n &\equiv& \frac{b \lambda \; r^n_H}{E} = 
\prod_j \left( 1 + \frac{l_j^2}{r_H^2} \right)^{- 1} 
\end{eqnarray}
so that $S$ can be written as 
\[
S = \rho \; S_0 \; \; , \; \; \; 
S_0 = \frac{4 \pi b}{n} (\lambda b)^{-\lambda}
(\sqrt{q} + \sqrt{\bar{q}}) E^\lambda
\]
where $S_0$ is the entropy in the non rotating case, namely
when $l_j = 0$.

One should next solve the equations obtained by setting to zero
the coefficients of $d q$, $d \bar{q}$, and $d E$ in the
expression $d \mathcal{F} = 0$. But it is much easier, and
equivalent, to use the results $l_j = \bar{l}_j$ and $r_H =
\bar{r}_H$ in the expression for the entropy $S$ and maximise
the resultant expression with respect to the remaining variables
$q$, $\bar{q}$, and $E$, subject to the two remaining
constraints $q - \bar{q} = Q$ and $b (q + \bar{q}) + 2E = M$.
Thus, using the Lagrange multipliers $B$ and $C$, we start from
the equation
\[
d \left\{ S + B (q - \bar{q} - Q) + C \left( b (q + \bar{q}) 
+ 2 E - M \right) \right\} = 0
\]
which implies that
\[
\left( \frac{\partial S}{\partial q} + B + b C \right) d q +
\left(\frac{\partial S}{\partial \bar{q}} - B + b C \right) 
d \bar{q} 
+ \left( \frac{\partial S}{\partial E} + 2 C \right) d E = 0 
\; .
\]
The coefficients of $d q$, $d \bar{q}$ and $d E$ must vanish. So
we get three equations. Eliminating $B$ and $C$ from them we get
\begin{eqnarray}\label{max:entropy}
\frac{\partial S}{\partial q} + \frac{\partial S}{\partial
\bar{q}} - b \frac{\partial S}{\partial E} = 0 \; .
\end{eqnarray}
Since $\frac{l_j}{r_H}=\frac{2 \pi J_j}{S}$ and $S = \rho S_0$,
equation (\ref{rho}) for $\rho$ can be written, following
\cite{ser-peet}, as an implicit function of $J_j$ and $S_0$ as
follows :
\begin{equation}\label{rhos0}
\rho^n = \prod_j \left( 1 + \left( \frac{2 \pi J_j}{\rho \; S_0}
\right)^2 \right)^{- 1} \; \; .
\end{equation}
Then, $\frac {\partial S}{\partial q}$ is given by
$\frac{\partial S}{\partial q} = \left(S_0 \frac{\partial
\rho}{\partial S_0} + \rho \right) \; \frac{\partial
S_0}{\partial q} \;$. The partial derivatives $\frac{\partial
S}{\partial \bar{q}}$ and $\frac{\partial S} {\partial E}$ can
be similarly expressed. Putting them back in equation
(\ref{max:entropy}), we find
\begin{eqnarray}\label{max:s_0}
\left(S_0 \frac{\partial \rho}{\partial S_0} + \rho \right)
\left\{ \frac{E^\lambda}{2 \sqrt{q}} + \frac{E^\lambda}
{2 \sqrt{\bar{q}}} - \lambda b (\sqrt{q} + \sqrt{\bar{q}})
E^{\lambda - 1} \right\} = 0 \; .
\end{eqnarray}
Using equation (\ref{rhos0}) for $\rho(S_0)$ one can show that
$S_0 \frac{\partial \rho}{\partial S_0} + \rho \ne 0$. So from
equation (\ref{max:s_0}) one then finds $E = 2 \; \lambda b \;
\sqrt{q \bar{q}} \;$.

Let $q = \frac{m}{2} e^{ 2 \theta}$ and $\bar{q} = \frac{m}{2}
e^{- 2 \theta}$. Then $E = \lambda b m$ and $M, Q, J_j$, and $S$
of the system are given, using equations (\ref{M_ft}) --
(\ref{S_ft}), by
\begin{eqnarray} 
M & = & b m \; \left( 2 \lambda + \cosh2\theta \right) 
\label{M1} \\
Q & = & m \; \sinh{2 \theta} \label{Q1} \\
J_j & = & \frac{4 b m}{n} \; \frac{l_j}{\sqrt2} \; 
\cosh\theta \label{J1} \\
S & = & \frac{8 \pi b m}{n} \; \frac{r_H}{\sqrt2} \; \cosh\theta
\label{S1} 
\end{eqnarray}
where $r_H$ is given implicitly by the equation 
$r_H^n \; \prod_j \left(1 + \frac{l_j^2}{r_H^2} \right) 
= m \;$. 

Now we compare the above expressions with the supergravity
ones. \footnote{Equations (\ref{M1}) -- (\ref{S1}) are obtained
from solving the variational equation $d \mathcal{F} = 0 \;$.
However, to show that the entropy $S$ is a maximum, one also has
to show that second order variations of $\mathcal{F}$ satisfy
appropriate conditions. This can be shown for simple cases but
the calculations become tedious for general cases. Hence,
following \cite{DGK} -- \cite{ks}, we assume in this paper that
the solutions obtained from solving $d \mathcal{F} = 0 \;$
corresponds to the maximum of entropy.} Setting $M = M_{sg}$, $Q
= Q_{sg}$, and $J_j = J_{j \; sg}$ gives $m = \mu$, $\theta =
\phi$, and $l_j = \sqrt{2} \; l_{j \; sg}$. We then have that
\begin{equation}
S(M, Q, J_j) \; = \; X \; S_{sg}(M, Q, J_j) \; \; , \; \; \;
X = \frac{1}{\sqrt{2}} \; \frac{r_H}{r_{H_{sg}}} \; \; .
\end{equation}
Thus, the field theoretic entropy $S$ differs from the
supergravity entropy $S_{sg}$ by a `deficit' factor $X$ given
above.

An implicit equation for $X$ can be obtained easily. Using
equation (\ref{sugra_r}) with $\mu = m$, $l_{j \; sg}^2 =
\frac{l_j^2}{2} \;$, and $r_{H_{sg}}^2 = \frac{r_H^2}{2 X^2}
\;$, and equation $r_H^n \; \prod_j \left(1 + \frac{l_j^2}
{r_H^2} \right) = m \;$ for $r_H$, it follows that $X$ is given
implicitly by the equation
\[
X^n = 2^{- \frac{n + 2}{2}} \prod_j 
\frac{r_H^2 + X^2 l_j^2}{r_H^2 + l_j^2} 
\]
and, hence, that the factor $X$ depends non trivially on $l_j$
and $r_H$, thus on black hole parameters $l_j$ and $m$. However,
in the non rotating case, $l_j = 0$ and the deficit factor $X$
reduces to just a numerical constant, namely $X = 2^{- \lambda}$
where $\lambda = \frac{n + 1}{2 n} \;$, see \cite{DGK, ser-peet,
bl}.

\subsection{Multicharge rotating black holes}

Multicharge black holes are described by stacks of intersecting
branes and antibranes. For $K$ charge black holes, we have $K$
types of brane antibrane pairs. There will be $2^K$ types of
gases, which can be understood as follows. The $K$ types of
branes, and antibranes, are numbered as $1, 2, \cdots, K$, and
$\bar{1}, \bar{2}, \cdots, \bar{K}$. K charge black holes can be
thought of as obtained by taking, for example, the stack $1,
\bar{2}, 3, \cdots, K$ and its anti stack $\bar{1}, 2, \bar{3},
\cdots, \bar{K}$, with a pair of gases on these two stacks.
Clearly, there are $2^{K - 1}$ such ways of constructing K
charge black holes and, consequently, the system is to be
thought of as containing a total of $2^K$ types of gases on
$2^K$ types of stacks \cite{ks}.

We assume, as in \cite{DGK} -- \cite{ks}, that each type of gas
has same energy $E$. This assumption is analogous to assuming
$E = \bar{E}$ in the single charge case, and is necessary to
obtain the entropy which matches that in supergravity
case. Then, the total mass $M$ and the charges $Q_i$, $i = 1, 2,
\cdots, K$, are given by
\begin{eqnarray}
M & = & b \sum_{i = 1}^K (q_i + \bar{q_i}) + 2^K E 
\label{M_mul} \\
Q_i & = & q_i - \bar{q_i}  \;\; . \label{Q_mul}
\end{eqnarray}

Each type of gas has different set of angular momentum
parameters, {\em i.e.} $I^{th}$ type of gas has parameters
$l_j^I$ where $I = 1, 2, 3, \cdots, 2^K$. For example, in single
charge case, $K = 1$ and there are two types of gases, with the
parameters denoted as $l_j^1 = l_j$ and $l_j^2 = \bar{l}_j$.
Since angular momenta and entropy of gases are assumed to be
additive, total angular momenta $J_j$ and total entropy $S$ will
be
\begin{eqnarray}
J_j & = & \frac{2 b}{n} \; \left( \frac{E}{\lambda b} 
\right)^{1 - \frac{K}{2}} \; \sum_{I = 1}^{2^K} l_j^I 
\prod_{i = 1}^K \left[ \sqrt{\tilde{q}_i} \right]^I
\label{J_mul} \\
S & = & \frac{4 \pi b}{n} \; \left( \frac{E}{\lambda b}
\right)^{1 - \frac{K}{2}} \; \sum_{I = 1}^{2^K} r_H^I 
\prod_{i = 1}^K \left[ \sqrt{\tilde{q}_i} \right]^I
\label{S_mul}
\end{eqnarray}
where $\lambda = \frac{n + 1}{n} - \frac{K}{2} \;$,
$\tilde{q}_i$ is either $q_i$ or $\bar{q}_i$ depending on the
stack, and $r_H^I$ which are now functions of $(E, \{ l^I_j \})$
are given by
\begin{equation}
(r^I_H)^n \prod_j \left( 1 + \frac{(l_j^I)^2}{(r^I_H)^2} \right)
= \frac{E}{\lambda b} \; .
\end{equation}

This is our field theoretic model for K charge rotating black
holes, with the entropy $S$ given by equation (\ref{S_mul}). The
parameters $q_i$, $\bar{q}_i$, $l_j^I$, and $E$ are arbitrary
subject only to the constraints that the quantities $M$, $Q_i$
and $J_j$ are fixed.

This system is dynamical because more and more brane antibrane
pairs can be created taking energy from massless gases, or
annihilated giving energy to them; moreover $l_j^I$ can flow
from one gas to another. The system reaches equilibrium in a
state where entropy is maximum for given $M$, $Q_i$, and
$J_j$. So, to obtain the equilibrium state, we maximise $S$
subject to the constraints that $M$, $Q_i$ and $J_j$ given in
equations (\ref{M_mul}) -- (\ref{J_mul}) are held fixed.

These constraints are incorporated by Lagrange multiplier
method. Hence we maximise the function $\mathcal{F} (q_i,
\bar{q}_i, l_j^I, E)$ defined by

\begin{eqnarray*}
\mathcal{F} & = & S + A_j \left( \sum_I \frac{2b}{n} l_j^I
\left( \frac{E} {\lambda b} \right)^{1 - \frac{K}{2}}
\sqrt{\left[ \prod \tilde{q}_i \right]^I } - J_j \right) \\
& + & \sum_i B_i (q_i - \bar{q}_i - Q_i) 
+ C \left( b \sum_i (q_i + \bar{q}_i) + 2^K E - M \right)
\end{eqnarray*}
where $A_j$, $B_i$ and $C$ are Lagrange multipliers. Varying
$\mathcal{F}$ with respect to $q_i$, $\bar{q}_i$, $l_j^I$, and
$E$, we have that $d \mathcal{F} = 0$ at the maximum. Hence,
\begin{eqnarray*}
& & 0 = \left(\frac{E}{\lambda b} \right)^{1 - \frac{K}{2}}
\sum_I  \sqrt{ \left[ \prod \tilde{q}_i \right]^I}
\left( \frac{\partial r_H^I}{\partial l_j^I} + \frac{A_j}{2 \pi}
\right) \; d l_j^I \\
& & + \sum_i{ \left( \frac{\partial S}{\partial q_i} 
+ A_j \frac{\partial }{\partial q_i} \sum_I \frac{2 b}{n} l_j^I
\left( \frac{E}{\lambda b} \right)^{1 - \frac{K}{2}} 
\sqrt{ \left[ \prod \tilde{q}_i \right]^I} + B_i + C b \right)
\; d q_i} \\
& & + \sum_i{ \left( \frac{\partial S}{\partial \bar{q}_i} 
+ A_j \frac{\partial}{\partial \bar{q}_i} \sum_I \frac{2 b}{n}
l_j^I \left( \frac{E}{\lambda b} \right)^{1 - \frac{K}{2}}
\sqrt{ \left[ \prod \tilde{q}_i \right]^I} - B_i + C b \right)
\; d \bar{q}_i} \\
& & + \left( \frac{\partial S}{\partial E} + A_j
\frac{\partial}{\partial E} \sum_I \frac{2 b}{n} l_j^I 
\left( \frac{E}{\lambda b} \right)^{1 - \frac{K}{2}}
\sqrt{ \left[ \prod \tilde{q}_i \right]^I } + 2^K C \right) 
\; dE 
\end{eqnarray*}
which implies that the coefficients of each $dl_j^I$, $dq_i$,
$d\bar{q}_i$ and $dE$ must vanish. Equating the coefficients of
each $l^I_j$, we get
\[
\frac{\partial r_H^I}{\partial l_j^I} = - \; \frac{A_j}{2\pi}  
\; \; \; \; \; \; \forall \; I  \; \; . 
\]
The functional dependence of $r_H^I$ on $l_j^I$ are same for
all $I$. So, just as in the single charge case, the above
$2^K$ equations always admit one solution where 
\[ 
l_j^1=l_j^2=l_j^3=\cdots=l_j^{2^K}\equiv l_j
\]
and, hence, $r_H^I \equiv r_H$ are also the same for all $I$.
Such a solution is always present and, as will be seen below, it
leads to the correct form of entropy. With this solution, the
angular momenta $J_j$ and entropy $S$ now become
\begin{eqnarray} 
J_j & = & \frac{2b}{n} \; l_j \; \left( \frac{E}{\lambda b} 
\right)^{ 1 - \frac{K}{2}} \prod_{i = 1}^K \; ( \sqrt{q_i} 
+ \sqrt{\bar{q_i}} ) \label{J_mull} \\
S & = & \frac{4 \pi b}{n} \; r_H \; \left( \frac{E}{\lambda b}
\right)^{1 - \frac{K}{2}} \prod_{i = 1}^K \; ( \sqrt{q_i} 
+ \sqrt{\bar{q_i}} ) \label{S_mull} 
\end{eqnarray}
where we have used the identity 
\[
\sum_{I = 1}^{2^K} \prod_{i = 1}^K 
\left[ \sqrt{\tilde{q}_i} \right]^I
= \prod_{i = 1}^K ( \sqrt{q_i} + \sqrt{\bar{q_i}} ) \; \; .
\]
Also, as in the single charge case, the entropy $S$ can be
written as
\[
S = \rho \; S_0 \; \; , \; \; \; 
S_0 = \frac{4 \pi b}{n}(b \lambda)^{- \lambda}
\prod_{i = 1}^K ( \sqrt{q_i} + \sqrt{\bar{q}_i} ) E^\lambda
\]
where $\rho$ is defined in equation (\ref{rho}) and $S_0$ is the
entropy in the non rotating case, namely when $l_j = 0$.

Procedding as in the single charge case, we use the above
results in the expression for the entropy $S$ and maximise the
resultant expression with respect to $q_i$, $\bar{q}_i$, and
$E$, subject to the constraints $q_i - \bar{q}_i = Q_i$ and
$\sum_i b (q_i + \bar{q}_i) + 2^K E = M$.  Thus, using the
Lagrange multipliers $B_i$ and $C$, we start from the equation
\[ 
d \left\{ S + \sum_i B_i (q_i - \bar{q}_i - Q_i) + C \left(
\sum_i b (q_i + \bar{q}_i) + 2^K E - M \right) \right\} = 0 
\]
which implies that 
\begin{eqnarray*}
\sum_i \left( \frac{\partial S} {\partial q_i} + B_i 
+ C b \right) d q_i
+ \sum_i \left( \frac{\partial S} {\partial \bar{q}_i} - B_i 
+ C b \right) d \bar{q}_i 
+ \left( \frac{\partial S} {\partial E} + 2^K C \right)d E = 0
\end{eqnarray*}
which, in turn, implies that 
\begin{eqnarray}\label{e.mul}
\frac{\partial S}{\partial q_i} + \frac{\partial S}
{\partial \bar{q}_i} - 2^{1 - K} b \frac{\partial S}
{\partial E} = 0 \; \; .
\end{eqnarray}
Using $S = \rho S_0$, with $\rho$ taken as a function of $S_0$,
one now obtains 
\[
\left( S_0 \frac{\partial \rho}{\partial S_0}
+ \rho \right) \left\{ \frac{E^\lambda}{2 \sqrt{q_i}}
+ \frac{E^\lambda}{2\sqrt{\bar{q}_i}} - 2^{1 - K} \lambda b 
E^{\lambda - 1} \right \} = 0 \; \; .
\]
One can show that $S_0 \frac{\partial \rho}{\partial S_0} + \rho
\ne 0$. It then follows that $2^K E = 4 \lambda b \; \sqrt{q_i
\; \bar{q}_i} \;$.

Let $q_i = \frac{m}{2} e^{2 \theta_i}$ and $\bar{q}_i =
\frac{m}{2} e^{ - 2 \theta_i} \;$. Then $E = 2^{1 - K} \lambda b
m$ and $M, Q_i, J_j$, and $S$ of the system are given, using
equations (\ref{M_mul}), (\ref{Q_mul}), (\ref{J_mull}), and
(\ref{S_mull}), by
\begin{eqnarray}
M & = & b m \; \left(2 \lambda + \sum_{i=1}^K \cosh 2 \theta_i
\right) \\
Q_i & = & m \; \sinh 2 \theta_i \\
J_j & = & 2^{\frac{K (K - 2)}{2}} \; \; \frac{4 b m}{n} \; 
l_j \; \prod_{i = 1}^K \cosh \theta_i \\
S & = & 2^{\frac{K (K - 2)}{2}} \; \; \frac{8 \pi b m}{n} \; 
r_H \; \prod_{i = 1}^K \cosh \theta_i
\end{eqnarray}
where $r_H$ is given implicitly by the equation $r_H^n \;
\prod_j \left(1 + \frac{l_j^2}{r_H^2} \right) = 2^{1 - K} \; m
\;$.

Now we compare the above expressions with the supergravity ones,
see footnote {\bf 1}. Setting $M = M_{sg}$, $Q_i = Q_{i \; sg}$,
and $J_j = J_{j \; sg}$ gives $m = \mu$, $\theta_i = \phi_i$,
and $l_j = 2^{\frac{K (2 - K)}{2}} \; l_{j \; sg}$. We then have
that
\begin{equation}
S(M, Q_i, J_j) \; = \; X \; S_{sg}(M, Q_i, J_j) \; \; , \; \; \;
X = 2^{\frac{K (K - 2)}{2}} \; \; 
\frac{r_H}{r_{H_{sg}}} \; \; . 
\end{equation}
Thus, the field theoretic entropy $S$ differs from the
supergravity entropy $S_{sg}$ by a `deficit' factor $X$ given
above.

An implicit equation for $X$ can be obtained easily. Using
equation (\ref{sugra_r}) with $\mu = m$, $l_{j \; sg}^2 
= 2^{K (K - 2)} \; l_j^2 \;$, and $r_{H_{sg}}^2 = 2^{K (K - 2)} \;
\frac{r_H^2}{X^2} \;$, and equation $r_H^n \; \prod_j \left(1 +
\frac{l_j^2} {r_H^2} \right) = 2 ^{1 - K} \; m \;$ for $r_H$, it
follows that $X$ is given implicitly by the equation 
\[
X^n = 2^{- \left( n + 1 - \frac{n K}{2} \right) K} \; 
\prod_j \frac{r_H^2 + X^2 l_j^2}{r_H^2 + l_j^2} 
\]
and, hence, that the factor $X$ depends non trivially on $l_j$
and $r_H$, thus on black hole parameters $l_j$ and $m$. However,
in the non rotating case, $l_j = 0$ and the deficit factor $X$
reduces to just a numerical constant, namely $X = 2^{- \lambda
K}$ where $\lambda = \frac{n + 1}{n} - \frac{K}{2}$, see
\cite{krama, ks}.

\section{The deficit factor $X$}

As seen above, the deficit factor $X$ is just a numerical
constant in the non rotating case. Then, the field theoretic
entropy differs from the supergravity one by just a numerical
factor. However, with rotation present, the entropies differ by
a factor which now depends on black hole parameters. If the
deficit factor is allowed to depend on the black hole parameters
then, very likely, any model can be argued to reproduce black
hole entropy upto such a deficit factor. The field theoretic
model of DGK then becomes less appealing.

In the field theoretic model of DGK, one can obtain $X = 1$ if
one assumes that each of the $2^K$ types of gases has an energy
$= 2^K E$. This has been shown in \cite{DGK, ght, ser-peet} for
the single charge case with or without rotation and, as will be
seen below, is true for the multicharge case also with or
without rotation.

Obviously, however, this assumption violates the conservation of
energy. As pointed out in \cite{ser-peet}, this violation is
perhaps due to the neglect of binding energies in the field
theoretic model. But the details of the binding energies are not
sufficiently well known and, hence, it is difficult at present
to verify this idea.

Note also that, in the field theoretic model, the energies of
the gases are to be assumed equal. This is unusual since, in
such systems, it is the temperatures which must be equal. No
physical mechanism is known which can enforce such an equality
of energies, instead of temperatures.

For the multicharge black holes with no rotation, it is shown in
\cite{krama, ks} that $X = 1$ can also be obtained if one
assumes that the available energy is all taken by one single gas
and, further, that all the brane tensions are reduced by a
factor of four. This single gas may perhaps be thought of as
some sort of superposition of $2^K$ possible types of gases, and
as ``living'' equally likely on any of the $2^K$ possible
stacks. Its entropy is an average of the entropies it has when
on these stacks. Clearly, in this approach, there is no
violation of conservation of energy. Also, the unusual
assumption that energies, not temperatures, of the $2^K$ types
of gases are equal becomes superfluous.

We will now show that $X = 1$ can be obtained similarly even
when rotation is present. The angular momenta of the single gas
is to be taken as an average of the angular momenta it has when
on $2^K$ possible stacks.

For this purpose, we introduce a set of parameters $\alpha,
\chi, \sigma$, and $\epsilon$ in the expressions for mass,
charges, angular momenta, and entropy as follows:
\begin{eqnarray}
M & = & \alpha \; b \sum_{i = 1}^K (q_i + \bar{q_i}) + 2^K E \;
= \; \alpha \; \sum_{i = 1}^K \tau_i V_i (N_i + \bar{N_i}) + 2^K
E \label{Malpha} \\
Q_i & = & \alpha \; (q_i - \bar{q_i}) \; = \; \frac{\alpha
\tau_i V_i}{b} \; (N_i - \bar{N_i}) \label{Qialpha} \\
J_j & = & \chi \; \frac{2 b}{n} \; l_j \; \left( 
\frac{\epsilon E} {\lambda b} \right)^{1 - \frac{K}{2}} \;
\prod_{i = 1}^K (\sqrt{q_i} + \sqrt{\bar{q_i}}) 
\label{ang.mom} \\
S & = & \sigma \; \frac{4 \pi b}{n} \; r_H \; \left( 
\frac{\epsilon E} {\lambda b} \right)^{1 - \frac{K}{2}} \;
\prod_{i = 1}^K (\sqrt{q_i} + \sqrt{\bar{q_i}}) \; \;. 
\label{sscale} 
\end{eqnarray}
where we have set $l_j^I = l_j$ and $r_H^I = r_H$ for $I = 1, 2,
3, \cdots, 2^K$, and used the relation between the numbers and
the charges of branes. The expression for $r_H$ is given by
\begin{eqnarray}
r_H^n \; \prod{ \left( 1 + \frac{l_j^2}{r_H^2} \right)} \; = \;
\frac{\epsilon E}{\lambda b} \; . 
\end{eqnarray}
Analysing this system as before, one finds $E = 2^{1 - K} \alpha
\lambda b m$ and
\begin{eqnarray*}
M & = & \alpha b m \; \left( 2 \lambda 
+ \sum_{i = 1}^K \cosh{2 \theta_i} \right) \\
Q_i & = &  \alpha  m \; \sinh{2\theta_i} 
\end{eqnarray*}
where we have set $q_i = \frac{m}{2} e^{2 \theta_i}$ and
$\bar{q}_i = \frac{m}{2} e^{ - 2 \theta_i} \;$. Setting $M =
M_{sg}$ and $Q_i = Q_{i \; sg}$ gives $\alpha m = \mu$ and
$\theta_i = \phi_i$. The angular momenta and entropy then
become
\begin{eqnarray*}
J_j & = & \frac{\chi}{\alpha^{\frac{K}{2}}} \; \left(
\frac{2^K}{\epsilon} \right)^{\frac{K - 2}{2}} \; \; 
\frac{4 b \mu}{n} \; l_j \; \prod_{i = 1}^K \cosh \theta_i \\
S & = & \frac{\sigma}{\alpha^{\frac{K}{2}}} \; \left(
\frac{2^K}{\epsilon} \right)^{\frac{K - 2}{2}} \; \; 
\frac{8 \pi b \mu}{n} \; r_H \; \prod_{i = 1}^K \cosh \theta_i
\end{eqnarray*}
where $r_H$ is given implicitly by the equation 
\begin{eqnarray}\label{normalised_R_H}
r_H^n \; \prod_j \left(1 + \frac{l_j^2}{r_H^2} \right) =
\epsilon \; 2^{1 - K} \; \mu \; .
\end{eqnarray}
Setting $J_j = J_{j \; sg}$ gives $l_j = \frac
{\alpha^{\frac{K}{2}}} {\chi} \; \left( \frac{2^K} {\epsilon}
\right)^{\frac{2 - K}{2}} \; l_{j \; sg} \;$. 
We then have that
\begin{equation}
S(M, Q_i, J_j) \; = \; X \; S_{sg}(M, Q_i, J_j) \; \; , \; \; \;
X = \frac{\sigma}{\alpha^{\frac{K}{2}}} \; 
\left(\frac{2^K}{\epsilon}\right)^{\frac{K - 2}{2}} \; 
\frac{r_H}{r_{H_{sg}}} \; \; . 
\end{equation}

Consider now the deficit factor $X$ given above. Let $\sigma =
\alpha = \chi = 1 \;$, which corresponds to the model of section
{\bf 3.2}. Then $\epsilon = 2^K$ will give $l_j = l_{j \; sg}$
and, hence, $r_H = r_{H_{sg}}$ as follows from equations
(\ref{normalised_R_H}) and (\ref{sugra_r}). One then obtains $X
= 1$. However, since $\epsilon = 2^K \;$, this means that energy
of each of the $2^K$ types of gases is $2^K E$. This method of
obtaining $X = 1$ is similar to that in \cite{DGK, ght,
ser-peet}, and violates conservation of energy. As we have just
shown, it is also applicable in the multicharge case with
rotation.

However, $X = 1$ can also be obtained if one chooses $\epsilon =
2^K$ and $\sigma = \chi = \alpha^{\frac{K}{2}}$. Then, again,
$r_H = r_{H_{sg}}$ as follows from equations
(\ref{normalised_R_H}) and (\ref{sugra_r}). If one further
chooses $\sigma = \frac{1}{2^K}$ and $\chi = \frac{1}{2^K}$ then
this choice of values for $\epsilon$, $\sigma$ and $\chi$ admits
the following interpretation: There is only a single gas which
may perhaps be thought of as some sort of superposition of $2^K$
possible types of gases. This single gas has all the available
energy $2^K E$, and ``lives'' equally likely on any of the $2^K$
possible stacks. Hence its entropy and angular momenta are the
averages of their $2^K$ possible values, as signified by the
choices $\sigma = \chi = \frac{1}{2^K}$, and equations
(\ref{ang.mom}) and (\ref{sscale}) for $J_j$ and $S$.

But this implies that $\alpha = \sigma^{\frac{2}{K}} = \frac{1}
{4}$. This may be taken to mean, see equations (\ref{Malpha})
and (\ref{Qialpha}), that brane tensions are effectively
normalised by this factor -- namely, they are all reduced by a
factor of four. This method of obtaining $X = 1$ is similar to
that in \cite{krama, ks}. We have now shown that it is valid
even when rotation is present. However, neither the details of
the superposition mentioned above nor the physical reason for
normalising brane tensions is clear to us at present.

\section{Conclusion}

We now summarise our results briefly and mention a few issues
that may be studied further.

We generalised the field theoretic model of DGK to multicharge
black holes with rotation. They are described as stacks of
intersecting branes and antibranes with $2^K$ types of gases on
them which are characterised by energies and angular momentum
parameters. Assuming the energies of the gases to be equal, and
following the methods of \cite{DGK, ght, ser-peet}, we found
that the angular momentum parameters for the gases must also be
equal. We found that the field theoretic and the supergravity
entropies are related by $S = X S_{sg}$ where, in the presence
of rotation, the deficit factor $X$ is not constant but depends
on black hole parameters.

The deficit factor $X$ not being constant makes the field
theoretic model less appealing. One can obtain $X = 1$ with a
further assumption which, however, violates conservation of
energy. This is perhaps due to the neglect of binding energies
but the details of the binding energies are not sufficiently
well known to verify this idea. Also, the assumption that the
energies, not temperatures, of the gases are equal is unusual.

We showed that $X = 1$ and, hence, $S = S_{sg}$ can be obtained,
as in \cite{krama, ks}, for rotating multicharge black holes
also. The physical interpretation of the field theoretic model
is also similar. In particular, the assumptions mentioned above
are superfluous and are not needed in this interpretation.

However, this interpretation involves a single gas, thought of
as some sort of superposition of $2^K$ types of gases, and also
a reduction of brane tensions by a factor of four. We do not
understand the nature of the superposition or the reason for the
reduction of brane tensions. It is important to understand these
aspects.

It is also important to understand if and how the field
theoretic description of non extremal black holes connects up
with the string/M theoretic description of extremal and near
extremal ones \cite{bhrev}. This may help in understanding the
Hawking radiation of non extremal blcak holes in terms of the
field theoretic models used here. See \cite{DGK, ser-peet} for
some discussions on these issues.

\end{document}